\documentclass[conference]{IEEEtran}
 \IEEEoverridecommandlockouts
\ifCLASSINFOpdf
\else
   \usepackage[dvips]{graphicx}
\fi
\usepackage{url}

\usepackage{cite}
\usepackage{graphicx,amsmath,amssymb,tikz,psfrag}
\usepackage{color}
\usepackage{fancyhdr} 
\usepackage{tikz}
\usetikzlibrary{shapes,arrows,chains}
\usepackage{xcolor}
\usepackage{algorithm}
\usepackage{algpseudocode}
\usepackage{enumerate}

\begin{document}

\title{ \huge The DoF Region of Order-$(K-1)$ Messages for the $K$-user MIMO Broadcast Channel with Delayed CSIT
 
}
\author{\IEEEauthorblockN{Tong Zhang\textsuperscript{$*$},  Shuai Wang\textsuperscript{$*$}, Taotao Wang\textsuperscript{$\dagger$}, and Rui Wang\textsuperscript{$*$}
	}%

	\IEEEauthorblockA{\textsuperscript{$*$}Department of Electrical and Electronic Engineering, Southern University of Science and Technology, Shenzhen, China }
 	\IEEEauthorblockA{\textsuperscript{$\dagger$}College of Electronics and Information Engineering, Shenzhen University, Shenzhen, China}
	Email:
zhangt7@sustech.edu.cn,
wangs3@sustech.edu.cn,
ttwang@szu.edu.cn,
wang.r@sustech.edu.cn.

\thanks{This work was supported by the National Natural Science Foundation of China under Grant 62001203, and the China Postdoctoral Science Foundation under Grant 2021M691453.}
}

\maketitle

\begin{abstract}
This paper theoretically characterizes the degrees-of-freedom (DoF) region of order-$(K-1)$ messages for the $K$-user multiple-input multiple-output (MIMO) broadcast channel with delayed channel state information at the transmitter (CSIT) and arbitrary antenna configurations, where the transmitter has $M$ antennas and the receiver $i=1,2,\cdots,K$ has $N_i$ antennas. For the converse, we first derive the DoF region of order-$(K-1)$ messages for the $K$-user MIMO broadcast channel with no CSIT and arbitrary antenna configurations with the aid of the proposed Genie-bound, and then establish the DoF outer region. For the achievability, we first design a 2-phase transmission scheme, and then propose a backward/forward cancellation algorithm for decoding. Specifically, we efficiently derive the achievable DoF region from the designed transmission scheme by transformation approach. The main implication of this paper is that for the order-$(K-1)$ messages of $K$-user MIMO broadcast channel, the DoF region with delayed CSIT is larger than the DoF region with no CSIT when $N_2<M$, where $N_1 \le N_2 \le \cdots \le N_K$.
\end{abstract}

 


\section{Introduction}

Wireless communications in the presence of mobility, such as vehicular communications, unmanned aerial vehicle (UAV) communications, and railway communications, have attracted many interests \cite{1,2,3,4,5,6,7,8}. While, one challenging problem is that the fast fading wireless channel aroused by the mobility may lead to a delay in the channel state information (CSI) feedback. 

If the channel state information at the transmitter (CSIT) is delayed, Maddah-Ali and Tse first studied the degrees-of-freedom (DoF) region for order-$j=1,2,\cdots,K$ messages of the $K$-user multiple-input-single-output (MISO) broadcast channel, where order-$j=1,2,\cdots,K$ messages are desired by $j$ receivers and the transmitter has $M$ antennas \cite{11}. Reference \cite{11} showed that for $K-j+1 \le M$ antenna configurations, the DoF regions were derived. Otherwise, there is a gap between achievable DoF region and DoF outer region. Following the line of \cite{11}, there are several research trends.  

One trend is to investigate how does the number of antennas impact on the DoF with delayed CSIT? For the two-user multiple-input multiple-output (MIMO)  broadcast channel with delayed CSIT and arbitrary antenna configurations, the DoF region was characterized in \cite{12}. For the three-user MIMO broadcast channel with delayed CSIT and symmetric antenna configurations (each receiver has $N$ antennas), for $M \le 2N$ and $3N \le M$ antenna configurations, the sum-DoF of order-$1$ messages was characterized in \cite{13}. While, for $2N < M < 3N$ antenna configurations, reference \cite{13} showed that there is a gap between the achievable sum-DoF and sum-DoF upper bound. In \cite{14}, for $2N < M < 3N$ antenna configurations, an improved achievable sum-DoF  of order-$1$ messages was derived based on a novel holistic design approach. For the three-user MIMO broadcast channel with delayed CSIT and arbitrary antenna configurations (receiver $i=1,2,3$ has $N_i$ antennas), for $M \le \max \{N_1+N_2, N_3\}$ antenna configurations, an achievable DoF region of order-$1$ messages and the sufficient condition to achieve the DoF region of order-$1$ messages were derived in \cite{15}. Recently, for $\max\{N_1+N_2, N_3\} < M$ antenna configurations, reference \cite{16} derived an achievable DoF region of order-$1$ messages by means of a novel transformation approach and the sufficient condition to achieve the DoF region of order-$1$ messages. In addition, under no CSIT or delayed CSIT, reference \cite{16} characterized the DoF region of order-$2$ messages for the three-user MIMO broadcast channel with arbitrary antenna configurations.

Another trend is to investigate how does the security constraint impact on the DoF with delayed CSIT? For the two-user MIMO broadcast channel with confidential messages  and delayed CSIT, the secure degrees-of-freedom (SDoF) region was characterized in \cite{21}. For the MIMO interference channel with confidential messages and delayed CSIT, an achievable sum-SDoF was derived in \cite{22}. For the MIMO interference channel with confidential messages and local output feedback, for $M\le N/2$, $M=N$, and $2N\le M$ antenna configurations, the sum-SDoF was characterized in \cite{23}. For the MIMO X channel with confidential messages, output feedback and delayed CSIT, the SDoF region was characterized in \cite{24}. For the MIMO X channel with confidential messages and delayed CSIT, an achievable sum-SDoF was obtained in \cite{25}. 

Moreover, there are trends on the alternating of instantaneous, delayed, and no CSIT (alternating CSIT) \cite{31}, the mix of perfect delayed CSIT and imperfect instantaneous (mixed CSIT) \cite{32}, and so on. 

In this paper, we investigate how the number of antennas impacts on the DoF region of order-$(K-1)$ messages for the $K$-user MIMO broadcast channel with delayed CSIT. In particular, we characterize the the DoF region of order-$(K-1)$ messages for the $K$-user MIMO broadcast channel with delayed CSIT and arbitrary antenna configurations. We generalize the converse and achievability proof approaches of the DoF region of order-$2$ messages for the three-user MIMO broadcast channel with delayed CSIT in \cite{16}. As for the converse in our settings, we first prove the DoF region of order-$(K-1)$ messages for the $K$-user MIMO broadcast channel with no CSIT, where we design a tight Genie-bound. Then, we establish the converse through the DoF region with no CSIT. As for the achievability in our settings, we design the transmission scheme, where the decoding is enabled by the design of  backward/forward cancellation algorithm. To concisely analyze the achievable DoF region of the proposed transmission scheme, we adopt the transformation approach introduced in \cite{16}. Our results show that, for order-$(K-1)$ messages of the $K$-user MIMO broadcast channel, utilizing delayed CSIT can be better than that by no CSIT when $N_2<M$, where $N_1 \le N_2 \le \cdots \le N_K$.

\textit{Notations}: The scalar, vector, and matrix are denoted by $a,\textbf{a}$, and $\textbf{A}$, respectively. The conjugate-transpose is denoted by $(\cdot)^H$. The higher order infinitesimal is denoted by $o(\cdot)$. The $\mathbb{R}_+^n$ denotes a $n$ non-negative real number tuple. The $\mathbb{C}^{n \times m}$ denotes the complex matrix with $n$ rows and $m$ columns. The  $\underline{\textbf{a}}$ or $\underline{\textbf{A}}$ is comprised of partial rows of $\textbf{a}$ or $\textbf{A}$. The block-diagonal matrix with blocks $\textbf{A}$ and $\textbf{B}$ is denoted by 
\begin{equation}
	\text{bd}\{\textbf{A}, \textbf{B}\} = 
	\begin{bmatrix}
		\textbf{A} & \textbf{0} \\
		\textbf{0} & \textbf{B}
	\end{bmatrix}. 
\end{equation}
 
\textit{Organizations}: The remaining of the paper is organized as follows: We introduce the system model in Section-II. Then, we present and discuss the main results (Theorems 1 and 2) of this paper in Section-III. The Theorem 1 is proven in Section-IV. The Theorem 2 is proven in Section-V. Finally, we draw our conclusions in Section-VI.

\section{System Model}  

\subsection{The $K$-user MIMO Broadcast Channel with Delayed CSIT and Arbitrary Antenna Configurations}

\begin{figure}[!ht]
	\begin{center}
		\begin{tikzpicture}[scale = 0.525]
		\node at (-1.5,2) {$T$};
		\node at (5.5,-1.5) {$R_K$};
		\node at (5.5,2)  {$R_2$};
		\node at (5.5,5)  {$R_1$};
		
		\node at (5.5,0.5) {$\vdots$};
		
		\draw [line width=0.65pt,-latex,dashed] (-1.25,7.5)--(-1.25,3.5);
		\draw [line width=0.65pt,dashed] (-1.25,7.5)--(0,7.5);
		\draw [line width=0.65pt,dashed] (4.25,7.5)--(5.5,7.5);
		\draw [line width=0.65pt,dashed] (5.5,7.5)--(5.5,6.2);
		
		\node at (2,7.5) {$\textbf{H}_i[t-\tau]$};
		\node at (2,6.8) {{\small{$i=1,\cdots,K$}}};
		\node at (2,4.5) {$\textbf{H}_1[t]$};
		\node at (2,2.5) {$\textbf{H}_2[t]$};
		\node at (2,-0.8) {$\textbf{H}_K[t]$};
		
		\draw [line width=0.65pt,-latex,dashed] (0,2)--  (4,-1.5);
		
		\draw [line width=0.65pt,-latex,dashed] (0,2)--  (4,2);
		
		\draw [line width=0.65pt,-latex,dashed] (0,2)--  (4,5);
		
		\draw [line width=0.5pt] (-1,2.7) -- (-0.5,2.7) -- (-0.3,3) -- (-0.3,2.4) -- (-0.5,2.7);
		\node at (-0.7,2.1) {$\vdots$};
		
		\node at (-0.7,0.5) {$M$};
		
		\draw [line width=0.5pt] (-1,1.3) -- (-0.5,1.3) -- (-0.3,1) -- (-0.3,1.6) -- (-0.5,1.3);

		\draw [line width=0.5pt]  (4.2,5.65) -- (4.5,5.45) -- (4.2,5.25) -- (4.2,5.65);
		\draw [line width=0.5pt]  (4.5,5.45) -- (5,5.45);
		\node at (4.6,5.1) {$\vdots$};
		\draw [line width=0.5pt]  (4.2,4.7) -- (4.5,4.5) -- (4.2,4.3) -- (4.2,4.7);
		\draw [line width=0.5pt]  (4.5,4.5) -- (5,4.5);
		

		\draw [line width=0.5pt]  (4.2,2.65) -- (4.5,2.45) -- (4.2,2.25) -- (4.2,2.65);
		\draw [line width=0.5pt]  (4.5,2.45) -- (5,2.45);
		\node at (4.6,2.1) {$\vdots$};
		\draw [line width=0.5pt]  (4.2,1.7) -- (4.5,1.5) -- (4.2,1.3) -- (4.2,1.7);
		\draw [line width=0.5pt]  (4.5,1.5) -- (5,1.5);
		
		\draw [line width=0.5pt]  (4.2,-0.85) -- (4.5,-1.05) -- (4.2,-1.25) -- (4.2,-0.85);
		\draw [line width=0.5pt]  (4.5,-1.05) -- (5,-1.05);
		\node at (4.6,-1.4) {$\vdots$};
		\draw [line width=0.5pt]  (4.2,-1.8) -- (4.5,-2) -- (4.2,-2.2) -- (4.2,-1.8);
		\draw [line width=0.5pt]  (4.5,-2) -- (5,-2);
		
		\node at (4.6,-2.5) {$N_K$};
		
		\node at (4.6,1) {$N_2$};
		
		\node at (4.6,4) {$N_1$};
		
		\draw [line width=0.55pt] (-2,1) rectangle (-1,3);
		\draw [line width=0.55pt,dashed] (-2.25,0) rectangle (0,3.5);

		\draw [line width=0.55pt,dashed]  (4,-3) rectangle (6.5,6.2);
		
		\draw [line width=0.55pt]  (5,-2.25) rectangle (6,-0.75);
		\draw [line width=0.55pt]  (5,4.25) rectangle (6,5.75);
		\draw [line width=0.55pt]  (5,1.25) rectangle (6,2.75);
		
		\draw [line width=0.55pt,dashed] (0,6.5) rectangle (4,8);
		
		\end{tikzpicture}
	\end{center}
	\caption{The $K$-user MIMO broadcast channel with delayed CSIT and arbitrary antenna configurations.} \label{C1Fig1}
\end{figure}
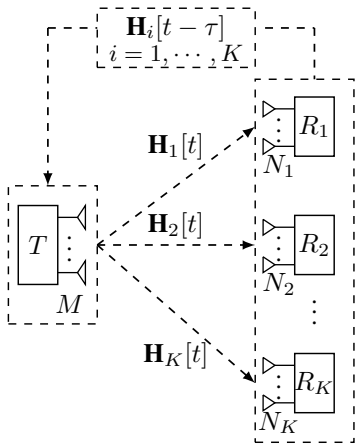

The $K$-user MIMO broadcast channel with delayed CSIT is depicted in Fig. 1, where the transmitter has $M$ antennas and receiver $i=1,2,\cdots,K$ has $N_i$ antennas. Without loss of generality, we have $N_1\le N_2 \le \cdots \le N_K$. \footnote{We denote the receiver with the most number of antennas by receiver $K$, and the receiver  with the least number of antennas by receiver $1$.} The communication process spans $n$ time slots. At the time slot $t = 1,2,\cdots,n$, the CSI matrix from the transmitter to the receiver $i=1,2,\cdots,K$ is denoted by $\textbf{H}_i[t] \in \mathbb{C}^{N_i \times M}$, whose elements are i.i.d. across space and time, and drawn from a continuous distribution. Therefore, we must have $\textbf{H}_i[t_1]$ is independent of $\textbf{H}_j[t_2]$ with probability of $1$, where $i \ne j$ and $t_1 \ne t_2$. Mathematically, the input-output relationship  for the receiver $i=1,2,\cdots,K$ at the time slot $t=1,2,\cdots,n$ is written as follows:
\begin{equation}  
\textbf{y}_i[t] = \textbf{H}_i[t]\textbf{x}[t] + \textbf{n}_i[t],
\end{equation}
where the transmit vector is denoted by $\textbf{x}[t]$ and the additive white Gaussian noise (AWGN) vector at receiver $i$ is $\textbf{n}_i[t] \sim {\cal{CN}}(0, \sigma^2)$. Given a maximal average transmit power $P$, $\textbf{x}[t]$ is subject to an average power constraint, i.e., $
\sum_{t=1}^n\textbf{x}[t]^H\textbf{x}[t] \le P
$. At the time slot $t$, due to the fast fading wireless channels, each receiver can only return the CSI matrices before time slot $t$ to the transmitter, as depicted in Fig. 1. The delayed CSI matrices are denoted by $\textbf{H}_i[t-\tau], 1 \le \tau$. Moreover, each receiver can obtain all the CSI matrices instantaneously.

\subsection{DoF Region of Order-$(K-1)$ Messages}

The order-$(K-1)$ message is referred to as the message desired by $K-1$ receivers simultaneously. Since the order-$(K-1)$ message is unwanted by one receiver, we denote all the order-$(K-1)$ messages by $W_{-1}, W_{-2}, \cdots,W_{-K}$, where the order-$(K-1)$ message $W_{-i}$ denotes the message desired by all receivers except the receiver $i$. 

We define the rate tuple $\{R_{-1}(P), \cdots,R_{-K}(P)\}$ of the order-$(K-1)$ messages as follows: There exists a code such that the probability of decoding error of the rate tuple approaches zero when the number of channel uses $n$ goes to infinity. The channel capacity region of the order-$(K-1)$ messages ${\cal{C}}_{K-1}(P)$ is defined as the region of all rate tuples satisfying the average power constraint. Mathematically, with definitions of rate tuple and channel capacity region, the DoF region of the order-$(K-1)$ messages is defined as follows: 
\begin{equation} 
 {\cal{D}}_{K-1}  =	
\left\{	
\begin{aligned}
 \{d_{-1}, d_{-2}, \cdots, d_{-K}\} \in \mathbb{R}^K_+|  \qquad \qquad \qquad \quad \,\,\,\, \\
 \{R_{-1}(P), R_{-2}(P), \cdots,R_{-K}(P)\} \in {\cal{C}}_{K-1}(P),   \\
 d_{-i} = \lim_{P \rightarrow {\cal{1}}} \frac{R_{-i}(P)}{\log_2 P}, \, i = 1,\cdots,K. \qquad \qquad \,\,
\end{aligned} 
\right\} 
\end{equation} 
It can be seen that the DoF region is the first-order approximation of channel capacity region. 

\section{Main Results and Discussions}

\textbf{Theorem 1 (DoF region with no CSIT)}: For the $K$-user MIMO broadcast channel with no CSIT and arbitrary antenna configurations, the DoF region of order-$(K-1)$ messages is given by 
\begin{equation} 
	\label{NoCSIT}
	{\cal{D}}_{K-1}^\text{N}  =	\left\{
	\begin{aligned}
		\{d_{-1}, d_{-2}, \cdots, d_{-K}\} \in \mathbb{R}^K_+| \qquad \quad  \\   
		\frac{\sum_{i=2}^K d_{-i}}{\min\{M,N_1\}} + \frac{d_{-1}}{\min\{M,  N_2\}} \le 1.
	\end{aligned} 
	\right\} 
\end{equation}
\begin{IEEEproof}
	Please refer to Section IV.
\end{IEEEproof}

\textbf{Theorem 2 (DoF region with delayed CSIT)}: For the $K$-user MIMO broadcast channel with delayed CSIT and arbitrary antenna configurations, the DoF region of order-$(K-1)$ messages is given by 
\begin{equation} 
	\label{DelayCSIT}
 	{\cal{D}}_{K-1}^\text{D}  =	\left\{
 	\begin{aligned}
 	 \{d_{-1}, d_{-2},\cdots, d_{-K}\} \in \mathbb{R}^K_+|  \qquad \qquad  \,\,\,\,\,\,\, \\   
	\frac{\sum_{i=2}^K d_{-i}}{\min\{M,N_1\}} + \frac{d_{-1}}{\min\{M,N_1 + N_2\}} \le 1, \\
\frac{\sum_{i=1,i \ne j}^K d_{-i}}{\min\{M,N_j\}} + \frac{d_{-j}}{\min\{M,N_1 + N_j\}} \le 1, \\
	 j=2,3,\cdots,K.  
	\end{aligned} 
	 \right\} 
	\end{equation}
\begin{IEEEproof}
 Please refer to Section V.
\end{IEEEproof}

\textbf{Remark (When delayed CSIT is useful)}: It can be seen from Theorems 1 and 2 that for order-$(K-1)$ messages of the $K$-user MIMO broadcast channel with arbitrary antenna configurations, ${\cal{D}}_{K-1}^\text{N} \subset {\cal{D}}_{K-1}^\text{D}$ holds, when $N_2 < M$. This implies that, for order-$(K-1)$ messages of the $K$-user MIMO broadcast channel, the delayed CSIT is useful when $N_2 < M$. 

\textbf{Remark (What if $N_1=N_2=\cdots=N_K=N$)}: When antenna configurations are symmetric, i.e., $N_1=N_2=\cdots=N_K=N$, the sum of strictly positive corner point  of DoF region of order-$(K-1)$ messages for the $K$-user MIMO broadcast channel with no CSIT can be derived from Theorem 1, which is given by $\min\{M,N\}$. Meanwhile, the sum of strictly positive corner point  of DoF region of order-$(K-1)$ messages for the $K$-user MIMO broadcast channel with delayed CSIT can be derived from Theorem 2, which is given by $K/((K-1)/\min\{M,N\}+1/\min\{M,2N\})$. We compare them in Fig. 2, where it implies that for fixed $M$ and $N$, adding a new receiver \textit{decreases} the sum-DoF of order-$(K-1)$ messages for the $K$-user MIMO broadcast channel with delayed CSIT. This is unlike the known result that adding a new receiver \textit{increases} the sum-DoF of order-$1$ messages for the $K$-user MIMO broadcast channel with delayed CSIT \cite{11}. This result can be verified in the designed achievable scheme for order-$K-1$ messages, where the coding gain in the Phase-II transmission becomes smaller when adding a new receiver. 

\begin{figure}
	\centering
	\includegraphics[width=3in]{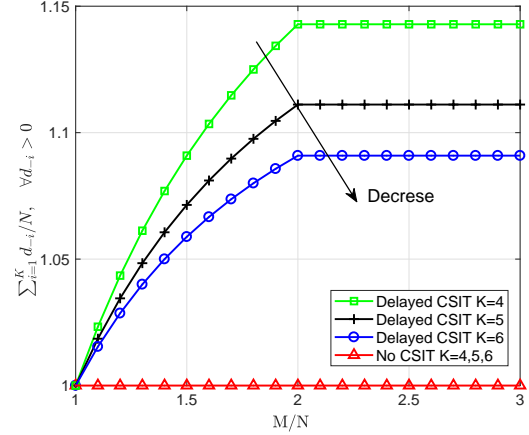}
	\caption{$\text{Sum-DoF}/N$ v.s. $M/N$}
\end{figure}

\section{Proof of Theorem 1}

\subsection{Converse Proof}

We propose to enhance the $K$-user MIMO broadcast channel with no CSIT by genie signalings. To be specific, the Genie provides $W_{-2},W_{-3},\cdots,W_{-K}$ to receiver 2. After that, according to the Fano's inequality and the setting of increasing order of decoding ability ($N_1 \le N_2 \le \cdots \le N_K$), we have the upper bounds of data rate, given by
\begin{subequations}
	\begin{eqnarray}
		n(\sum_{i=2}^K R_{-i}) \le I(W_{-2},\cdots,W_{-K};\textbf{y}_1^n|\textbf{H}^n) + o(\log \text{SNR}), \label{A1}\\
		nR_{-1} \le I(W_{-1};\textbf{y}_2^n|\textbf{H}^n,W_{-2},\cdots,W_{-K}) + o(\log \text{SNR}), \label{A2}
	\end{eqnarray}
\end{subequations}
where the collection of output signals at receiver 1 across $n$ channel uses is denoted by $\textbf{y}_1^n$, and the collection of all CSI matrices across $n$ channel uses is denoted by $\textbf{H}^n$. Based on \eqref{A1} and \eqref{A2}, we have \eqref{B}, shown at the top of next page,
\begin{figure*}
	\begin{eqnarray}
		&&\frac{n(\sum_{i=2}^K R_{-i})}{\min\{M,N_1\}} + \frac{nR_{-1}}{\min\{M,N_2\}} \nonumber 
		\\ 
		&& \le \frac{I(W_{-2},\cdots,W_{-K};\textbf{y}_1^n|\textbf{H}^n)}{\min\{M,N_1\}} 
  + \frac{I(W_{-1};\textbf{y}_2^n|\textbf{H}^n,W_{-2},\cdots,W_{-K})}{\min\{M,N_2\}}   + o(\log \text{SNR}) \nonumber 
		\\ 
		&& = \frac{h(\textbf{y}_1^n|\textbf{H}^n)-h(\textbf{y}_1^n|\textbf{H}^n,W_{-2},\cdots,W_{-K})}{\min\{M,N_1\}}   
		+  \frac{h(\textbf{y}_2^n|\textbf{H}^n,W_{-2},\cdots,W_{-K})-h(\textbf{y}_2^n|\textbf{H}^n,W_{-1},\cdots,W_{-K})}{\min\{M,N_2\}}  
	  + o(\log \text{SNR}) \nonumber 
		\\ 
		&&  \overset{(a)}{=} \frac{h(\textbf{y}_1^n|\textbf{H}^n)}{\min\{M,N_1\}} 
		+  \frac{h(\textbf{y}_2^n|\textbf{H}^n,W_{-2},\cdots,W_{-K})}{\min\{M,N_2\}} 	- \frac{h(\textbf{y}_1^n|\textbf{H}^n,W_{-2},\cdots,W_{-K})}{\min\{M,N_1\}} 
		+ o(\log \text{SNR})\nonumber 
		\\ 
		&&   \overset{(b)}{\le} \frac{h(\textbf{y}_1^n|\textbf{H}^n)}{\min\{M,N_1\}} +  o(\log \text{SNR}) \nonumber \\
		&& \overset{(c)}{\le} n \log \text{SNR} + o(\log \text{SNR}) 	\label{B}
	\end{eqnarray}
\hrule
\end{figure*}
where the reasons of critical steps are explained as follows:
\begin{enumerate}[(a)]
	\item is from $h(\textbf{y}_2^n|\textbf{H}^n,W_{-1},\cdots,W_{-K}) = o(\log \text{SNR})$
	\item is from  $\frac{h(\textbf{y}_2^n|\textbf{H}^n,W_{-2},\cdots,W_{-K})}{\min\{M,N_2\}} 	- \frac{h(\textbf{y}_1^n|\textbf{H}^n,W_{-2},\cdots,W_{-K})}{\min\{M,N_1\}} \le 0$ by applying \cite[Lemma and (18)]{41}
	\item is from $h(\textbf{y}_1^n|\textbf{H}^n) \le \min\{M,N_1\} n\log \text{SNR}$
\end{enumerate}
Finally, we can obtain the converse, having the same expression as \eqref{NoCSIT}, by dividing both sides of \eqref{B} by $n \log \text{SNR}$ and taking the limit of $n$. This completes the proof.

\subsection{Achievability Proof}

To achieve corner points of \eqref{NoCSIT}, we transmit $\min\{M, N_1\}$ order-$(K-1)$ symbols for $W_{-i},i=2,3,\cdots,K$, or $\min \{M, N_2\}$ order-2 symbols for $W_{-1}$, in one TS. The entire region can be attained through time-sharing of the schemes used in achieving the corner points. This completes the proof.

\section{Proof of Theorem 2}

\subsection{Converse Proof}

 The $K$-user MIMO broadcast channel with delayed CSIT is enhanced to the physically degraded broadcast channel, by providing the output of receiver $i$ to receivers $i+1, \cdots, K$. According to \cite{41,42}, the output feedback (delayed feedback) will not change the capacity region of the physically degraded broadcast channel with no CSIT. Thus, the DoF region of order-$(K-1)$ messages for the physically degraded MIMO broadcast channel with delayed CSIT is equal to the DoF region of order-$(K-1)$ messages for the MIMO broadcast channel with no CSIT. Applying Theorem 1 and exhaustively permuting all receiver indexes, we can express the following DoF outer region:
\begin{equation} 
	 {\cal{D}}_{K-1}^\text{D-Outer} = \left\{\begin{aligned} 
	(d_{-1}, \cdots , d_{-K}) \in \mathbb{R}^K_+|  \qquad \qquad \qquad \quad \,\,\,\,\,\,  \\
	\frac{\sum_{i=2}^K d_{-i}}{\min\{M,N_1\}} + \frac{d_{-1}}{\min\{M,N_1 + N_i\}} \le 1, \,\,\,\,\, \\
	i=2,3,\cdots,K, \\
	\frac{\sum_{i=1,i\ne 2}^K d_{-i}}{\min\{M,N_2\}} + \frac{d_{-2}}{\min\{M,N_2 + N_i\}} \le 1, \,\,\,\, \\
i=1,3,4,\cdots,K, \\
\vdots \qquad \qquad \qquad   \\
	\frac{\sum_{i=1,i\ne K}^K d_{-i}}{\min\{M,N_K\}} + \frac{d_{-K}}{\min\{M,N_K + N_i\}} \le 1, \\
i=1,2,\cdots,K-1. 
	\end{aligned}
	\right\} \label{DoFregion1}
	\end{equation}
Notice that the DoF outer region in \eqref{DoFregion1} is not concisely expressed. For example, the following $K-1$ inequalities:
\begin{eqnarray}
	\frac{\sum_{i=2}^K d_{-i}}{\min\{M,N_1\}} + \frac{d_{-1}}{\min\{M,N_1 + N_i\}} \le 1, \nonumber \\
i=2,3,\cdots,K,  
	\end{eqnarray}
are dominated by the first inequality, given by
\begin{equation}
	\frac{\sum_{i=2}^K d_{-i}}{\min\{M,N_1\}} + \frac{d_{-1}}{\min\{M,N_1 + N_2\}} \le 1. 
	\end{equation}
After eliminating all the redundant inequalities, we have the same expression of \eqref{DelayCSIT}, which completes the proof.

\subsection{Achievability Proof}

If $M \le N_2$, it can be verified that ${\cal{D}}_{K-1}^\text{D} = {\cal{D}}_{K-1}^\text{N}$. Hence, we focus on $N_2 < M$ Case. 
In this case, we design a transmission scheme, and show that the proposed transmission scheme can attain the converse of DoF region.

The proof has three steps and is sketched as follows: 1) We transmit one kind of undetermined number of order-$(K-1)$ symbols in each TS. 2) Using delayed CSIT, we re-construct the interference for order-$(K-1)$ symbol transmission
and transmit the coded interference, which is desired by $K$ receivers and referred to as order-$K$ symbol. 3) Through transformation approach, we analyze the achievable DoF region of the transmission scheme, and determine the number of transmit antennas for order-$(K-1)$ symbol transmission. We elaborate on the proposed transmission scheme as below.  

\textit{Phase-I (Order-$(K-1)$ Symbol Transmission)}: This phase aims at transmitted all the order-$(K-1)$ symbols encoded from order-$(K-1)$ messages. This phase spans $\sum_{i=1}^K T_{i}$ time slots, where the duration of order-$(K-1)$ symbol (encoded from $W_{-i}$) transmission is denoted by $T_{i}$. For ease of expression, we define the following holistic CSI matrices:
\begin{eqnarray}
	&& \textbf{H}_{i}^{\text{Ph-I},1} \triangleq \text{bd}\left\{\textbf{H}_i[1],\cdots,\textbf{H}_i[T_{1}]\right\},  \nonumber \\
	&& \textbf{H}_{i}^{\text{Ph-I},2} \triangleq \text{bd}\left\{\textbf{H}_i[T_{1}+1],\cdots,\textbf{H}_i[T_{1}+T_{2}]\right\}, \nonumber \\
	&& \qquad \qquad \vdots \nonumber \\
	&& \textbf{H}_{i}^{\text{Ph-I},K} \triangleq \text{bd}\left\{\textbf{H}_i[\sum_{i=1}^{K-1} T_{i} + 1],\cdots,\textbf{H}_i[\sum_{i=1}^{K} T_{i}]\right\}, \nonumber 
\end{eqnarray}
where $i = 1,2,\cdots,K$. Note that for DoF analysis, the impact of AWGN is omitted for simplicity.

From time slot $1$ to time slot $T_1$, the message $W_{-1}$ is encoded into $Q_{1}T_{1}$ symbols, denoted by $\textbf{s}_{-1} \in \mathbb{C}^{Q_{1}T_{1}}$. To transmit $\textbf{s}_{-1}$, the transmitter sends $Q_{1}$ symbols with $Q_{1}$ antennas, where $Q_1$ will be determined later on. Accordingly, the holistic received signals for receiver $i=1,2,\cdots,K$ over time slots $1,2,\cdots,T_1$, are given by
\begin{equation}
	\textbf{y}_i^{\text{Ph-I},1} = \textbf{H}_{i}^{\text{Ph-I},1}\textbf{s}_{-1}, \quad i = 1,2,\cdots,K. 
\end{equation}

From time slot $T_1+1$ to time slot $T_1+T_2$, the message $W_{-2}$ is encoded into $Q_2T_2$ symbols, denoted by $\textbf{s}_{-2} \in \mathbb{C}^{Q_2T_2}$. To transmit $\textbf{s}_{-2}$, the transmitter sends $Q_{2}$ symbols with $Q_{2}$ antennas, where $Q_2$ will be determined later on. Accordingly, the holistic received signals for receiver $i=1,2,\cdots,K$ over time slots 
$T_1+1,\cdots,T_1+T_2$, are given by
\begin{equation}
	\textbf{y}_i^{\text{Ph-I},2} = \textbf{H}_{i}^{\text{Ph-I},2}\textbf{s}_{-2}, \quad i = 1,2,\cdots,K. 
\end{equation}

From time slot $\sum_{k=1}^{j-1}T_{k}+1$ to time slot $\sum_{k=1}^j T_{k}$, the message $W_{-j},j=3,4,\cdots,K$, is encoded into $Q_{j}T_{j}$ symbols, denoted by $\textbf{s}_{-j} \in \mathbb{C}^{Q_{j}T_{j}}$. To transmit $\textbf{s}_{-j},j=3,4,\cdots,K$, the transmitter sends $Q_{j}$ symbols with $Q_{j}$ antennas, where $Q_j$ will be determined later on.  Accordingly, the holistic received signals for receiver $i=1,2,\cdots,K$ over time slots $\sum_{k=1}^{j-1}T_{k}+1, \cdots, \sum_{k=1}^j T_{k}$, are given by
\begin{equation}
	\textbf{y}_i^{\text{Ph-I},j} = \textbf{H}_{i}^{\text{Ph-I},j}\textbf{s}_{-j}, \quad i = 1,2,\cdots,K. 
\end{equation}
where $j=3,4,\cdots,K$. We enforce $Q_1,Q_2,\cdots,Q_K > N_1$. Therefore, all the transmitted symbols cannot be immediately decoded at all receivers. 

\textit{Phase-II (Order-$K$ Symbol Transmission)}: This phase spans $T$ TSs. At the beginning of the Phase-II, all the CSI matrices of Phase-I are fed back to the transmitter.  Using these CSI matrices, the transmitter can re-construct $\textbf{y}_1^{\text{I},1}$, $\textbf{y}_2^{\text{I},2}$, $\cdots$, $\textbf{y}_K^{\text{I},K}$, where each of them is interference for only one receiver. For example, $\textbf{y}_1^{\textbf{I},1}$ is the interference for receiver 1, but desired by the other receivers, since it contains the message $W_{-1}$. Hence, $\textbf{y}_1^{\textbf{I},1}$ supplements the lacking equations for decoding the message $W_{-1}$ at receivers $2,3,\cdots,K$. 

In order to deliver $\textbf{y}_1^{\text{I},1}$, $\textbf{y}_2^{\text{I},2}$, $\cdots$, $\textbf{y}_K^{\text{I},K}$ to target receiver to assist the decoding of order-$(K-1)$ symbols, we encode them into order-$K$ symbols, which is designed as follows:	 
\begin{equation}
	\textbf{x}_{\text{order-K}} \triangleq \begin{bmatrix}
		\underline{\textbf{y}}_{1}^{\text{I},-1} + \underline{\textbf{y}}_{2}^{\text{I},-2} \\
		\underline{\textbf{y}}_{2}^{\text{I},-2} + \underline{\textbf{y}}_{3}^{\text{I},-3} \\
		\underline{\textbf{y}}_{3}^{\text{I},-3} +   \underline{\textbf{y}}_{4}^{\text{I},-4} \\
		\cdots \\
		\underline{\textbf{y}}_{K-1}^{\text{I},-(K-1)} + \underline{\textbf{y}}_{K}^{\text{I},-K}
	\end{bmatrix} \in \mathbb{C}^{BT},
\end{equation}
where $\underline{\textbf{y}}_{i}^{\text{I},i} ,i = 1,2,\cdots,K$ is truncated from  $\textbf{y}_{i}^{\text{I},i}$. Note that $\textbf{x}_{\text{order-K}}$ are desired by all receivers. The transmission of $\textbf{x}_{\text{order-K}}$ is designed by the time division manner and with $B$ transmit antennas and $T$ time slots, so that all order-$K$ symbols can be immediately decoded at each receiver. In particular, $B$ can be given as follows: If $N_p \le M < N_{p+1}$, then $B = N_p$.

 Once upon the receiver decodes $\textbf{x}_{\text{order-K}}$, each receiver can retrieve the desired equations by the following backward/forward cancellation algorithm (Algorithm 1), whose idea is to iteratively acquire the desired equations by cancellation the known signals from the order-$K$ symbols. 
\begin{algorithm}[!t]
	\caption{Backward/Forward Cancellation Algorithm}  
	\hspace*{0.02in} {\bf Input:} \hspace*{0.02in} 
	the index of receiver $i$, $\underline{\textbf{y}}_{i}^{\text{I},i}$\\
	\hspace*{0.02in} {\bf Output:}  $\underline{\textbf{y}}_{1}^{\text{I},1},\underline{\textbf{y}}_{2}^{\text{I},2},\cdots,\underline{\textbf{y}}_{K}^{\text{I},K}$ except $\underline{\textbf{y}}_{i}^{\text{I},i}$
	\begin{algorithmic}[1]
		\State Set $t = \underline{\textbf{y}}_{i}^{\text{I},i}$
		\State For $j=i:-1:2$ (\textit{Backward cancellation})
		\State $\qquad \underline{\textbf{y}}_{j-1}^{\text{I},j-1} =  \underline{\textbf{y}}_{j-1}^{\text{I},j-1} +  \underline{\textbf{y}}_{j}^{\text{I},j} - t $
		\State $\qquad t = \underline{\textbf{y}}_{j-1}^{\text{I},j-1}$
		\State Set $t = \underline{\textbf{y}}_{i}^{\text{I},i}$
		\State For $j=i:1:K-1$ (\textit{Forward cancellation})  
		\State  $\qquad \underline{\textbf{y}}_{j+1}^{\text{I},j+1} =  \underline{\textbf{y}}_{j+1}^{\text{I},j+1} +  \underline{\textbf{y}}_{j}^{\text{I},j} - t$
		\State $\qquad t = \underline{\textbf{y}}_{j+1}^{\text{I},j+1}$	 
	\end{algorithmic}
\end{algorithm}

\textit{Achievable DoF Region Analysis by Transformation Approach \cite{16}}: In the following, we analyze the achievable DoF region of the proposed transmission scheme, where the aim is to determine the number of transmit antennas $Q_1,Q_2,\cdots,Q_K$ to  attain the converse of DoF region. The idea of transformation approach is to transform the decoding condition into the achievable DoF region, through the achievable DoF tuple expression. Thus, this approach avoids the complicate verification that the corner points of the DoF outer region are achievable. The analysis of achievable DoF region is elaborated below.

To begin with, according to the transmission scheme, the decoding condition of the  scheme can be stated as follows:
\begin{subequations}
	\begin{eqnarray}
		\sum_{i=2}^K T_{i}(Q_i - N_1) \le TN_1, \label{E4.1.1}\\
		\sum_{i=1,i \ne 2}^K T_{i}(Q_i - N_2) \le TN_2, \label{E4.1.2}\\
		\vdots \qquad \qquad \qquad \nonumber \\
		\sum_{i=1}^{K-1} T_{i}(Q_i - N_K) \le TN_K, \label{E4.1.3}
	\end{eqnarray}
\end{subequations}
where \eqref{E4.1.1}-\eqref{E4.1.3} represent that the number of lacking equations at receiver $i=1,2,\cdots,K$ can be compensated by the equations provided by Phase-II. 
For simplicity, we denote $\beta  = \sum_{i=1}^K T_i + T$. Adding $(\beta  - T)N_1$, $(\beta  - T)N_2$,$\cdots$,$(\beta  - T)N_K$ at both sides of inequalities \eqref{E4.1.1}-\eqref{E4.1.3}, respectively, we have
\begin{subequations}
	\begin{eqnarray}
		\sum_{i=2}^K T_{i}Q_i + T_1N_1 \le \beta N_1, \label{E4.2.1}\\
		\sum_{i=1,i \ne 2}^K T_{i}Q_i + T_2N_2 \le \beta N_2, \label{E4.2.2}\\
		\vdots \qquad \qquad \qquad \nonumber \\
		\sum_{i=1}^{K-1} T_{i}Q_i + T_{K}N_K \le \beta N_K. \label{E4.2.3}
	\end{eqnarray}
\end{subequations}
Next, dividing both sides of the inequalities \eqref{E4.2.1}, \eqref{E4.2.2}, $\cdots$, \eqref{E4.2.3}  by $\beta N_1$, $\beta N_2$, $\cdots$, $\beta N_K$, respectively, we have
\begin{subequations}
	\begin{eqnarray}
		\sum_{i=2}^K \frac{T_{i}Q_i}{\beta} \frac{1}{N_1} + \frac{T_{1}Q_1}{\beta}\frac{1}{Q_1} \le 1, \label{E4.3.1}\\
		\sum_{i=1,i \ne 2}^K \frac{T_{i}Q_i}{\beta} \frac{1}{N_2} + \frac{T_{2}Q_2}{\beta}\frac{1}{Q_2} \le 1, \label{E4.3.2}\\
		\vdots \qquad \qquad \qquad \nonumber \\
		\sum_{i=1}^{K-1} \frac{T_{i}Q_i}{\beta} \frac{1}{N_K} + \frac{T_{K}Q_K}{\beta}\frac{1}{Q_K} \le 1. \label{E4.3.3}
	\end{eqnarray}
\end{subequations}
The achievable DoF tuple is expressed as 
\begin{equation}
	(d_{-1},d_{-2},\cdots,d_{-K}) = \left( \frac{T_{1}Q_1}{\beta}, \frac{T_{2}Q_2}{\beta}, \cdots, \frac{T_{K}Q_K}{\beta} \right). \label{tuple} 
\end{equation}
Therefore, after setting 
\begin{subequations}
\begin{eqnarray}
&& Q_1 =  \min\{M, N_1+N_2\} \\
&& Q_j= \min\{M,N_1+N_j\}, j=2,3,\cdots,K.
\end{eqnarray}	
\end{subequations}
It can be verified that the derived achievable DoF region is equal to the converse of DoF region, where the achievable DoF region is derived by substituting the \eqref{tuple} into \eqref{E4.3.1}-\eqref{E4.3.3}. 

\section{Conclusions}
 
We have characterized the DoF region of order-$(K-1)$ messages for the $K$-user MIMO
broadcast channel with delayed CSIT and arbitrary antenna configurations. Moreover, we have derived the DoF region of order-$(K-1)$ messages for the $K$-user MIMO
broadcast channel with no CSIT and arbitrary antenna configurations. Therefore, we have answered the question that when using delayed CSIT is better than using no CSIT in terms of DoF w.r.t. antenna configurations.


 
\bibliographystyle{IEEEtran}
\bibliography{ICCCRef}
 
\end{document}